\documentclass[journal,10pt]{IEEEtran}
\usepackage{mathtools}
\usepackage{amsmath,color,amsthm}
\usepackage{breqn}
\usepackage{cite}
\usepackage{algorithm,algpseudocode}
\usepackage{caption}
\usepackage{lipsum}
\usepackage{cuted}
\usepackage{enumerate}
\usepackage{graphicx}
\usepackage{float}
\usepackage{mathrsfs}

\usepackage{amsmath,amsthm,amssymb}

\usepackage{wasysym} 

\DeclarePairedDelimiterX\Basics[1](){ #1}

\date{\today}

\begin{document}
	\title{Large-Scale NOMA: Promises for Massive Machine-Type Communication} 
\author{Ekram Hossain and Yasser Al-Eryani\thanks{The authors are with the Department of Electrical and Computer Engineering at the University of Manitoba, Canada (Emails: \{Ekram.Hossain, Yasser.Aleryani\}@umanitoba.ca).}}
 	\maketitle
	\begin{abstract}
We investigate on large-scale deployment of non-orthogonal multiple access (NOMA)  for improved spectral and power efficiency in cellular networks to provide massive wireless connectivity (e.g. for machine-type communication [mMTC]). 
First, we describe the basics of single-antenna NOMA technology and its extension to co-located multiple-antenna NOMA as well as coordinated multipoint transmission (CoMP)-enabled NOMA technologies. 
Then we discuss some of the practical challenges of large-scale deployment of NOMA such as the inter-NOMA-interference (INI), inter-cell interference, and hardware implementation complexity.
To this end, we present one key enabling technique to overcome the challenges  of large-scale deployment of NOMA.
Generally speaking, for a feasible large-scale NOMA implementation, sophisticated diversity enhancing techniques can be used to compensate for the degradation in coding gain and to decrease the complexity resulting from excessive INI and increased level of required successive interference cancellation (SIC).  
Furthermore, to massively extend NOMA over the network coverage area, NOMA transmitters have to cooperate in a generalized manner to serve all nearby users simultaneously.
	\end{abstract}
\section{Introduction}
Supporting a massive numbers of wireless-connected devices that communicate with each other simultaneously will be an inherent property of the fifth generation (5G) and beyond 5G (B5G) systems.
Massive-scale connectivity between various communication devices (e.g. vehicles, sensors, mobiles, etc.) will enable us to form the internet of everything (IoE) that bridges the gap between the `Cyber' and the `physical' worlds.
For this so called massive machine-type communication (mMTC), efficient, reliable, and fast communication methods will be required to enhance and control data flows  among these devices. 
As an example, autonomous driving and intelligent transportation systems require that hundreds of machines to be coordinating and communicating simultaneously within a certain geographical area.
Communications among these machines  must be reliable and fast in order to provide a safe driving environment.  This can be achieved by increasing the capacity of wireless systems to support significantly higher data  rates and/or number of devices. 
It may also be used to reduce power consumption on mMTC distributed devices.


The capacity of wireless systems however is limited by the availability of spectrum resources. To improve the capacity under limited spectrum resources,  sophisticated designs for wireless access and resource allocation will be required. In this context, 
non-orthogonal multiple-access (NOMA) has emerged as a potential technology to tackle the problem of improving wireless network capacity under spectrum scarcity \cite{NOMA2006}.
The main concept of NOMA (in uplink and downlink) is: in a certain frequency sub-band, signals for multiple users are superimposed in the power domain such that the resultant received signal has a distinct power levels for every user. 
At the receiver side, successive interference cancellation (SIC) is used to filter out the undesired signals.
Generally, SIC can be conducted at the codeword level, symbol level or using reduced complexity-maximum likelihood detection (R-ML). However, codeword and symbol levels SIC are preferable methods due to their ability to suppress interference from adjacent devices and their relatively low complexity   \cite[Table 2]{NOMASIC2018}.

Theoretically, NOMA has been observed to outperform orthogonal multiple access  (OMA) schemes such as OFDMA, CDMA and TDMA, in terms of spectral efficiency.
Additionally, under the low-rate requirements of mMTC terminal devices, large-scale NOMA can achieve a significantly high power efficiency when a similar bandwidth as that of OMA system is utilized\footnote{In this article, large-scale NOMA refers to the use of relatively high number of users per single NOMA cluster throughout a single frequency sub-band.}.
However, two major drawbacks of NOMA on its simplest form (single antenna and single base station at transmitter(s) and receiver(s))  limit its usage in a large-scale regime. 
The first drawback is the increased level of inter-NOMA-interference (INI), which is the interference caused by other users using the same NOMA cluster. 
When the number of NOMA users in a certain NOMA cluster increases, those users with lower link quality will have to deal with larger number of unfiltered INI signals.
The second drawback is the hardware complexity of SIC  that increases significantly with the number of users per NOMA cluster \cite{NCOMASIC2017}. 

We investigate on a number of enabling techniques that may be used to overcome the challenges of large-scale NOMA (e.g. high INI and high hardware complexity requirements).
NOMA can be deployed either in the uplink or the downlink of a wireless system. 
In this article, we focus on downlink NOMA as discussions can be easily applied to uplink NOMA.
The rest of this article is organized as follows. Section II describes the basics of  single-antenna, multi-antenna, and CoMP-enabled NOMA technologies.
The hardware requirements for large-scale NOMA are discussed in Section III. Section IV outlines a key enabling technique that can support implementation of large-scale NOMA. 
We conclude the article in Section V.  
\section{NOMA Techniques}
This section describes several enabling technologies the NOMA scheme can be integrated with.
\subsection{Single-Input Single-Output (SISO)-NOMA}
\begin{figure}[!htb]
		\centering		\includegraphics[height=5.8cm, width=7.75cm]{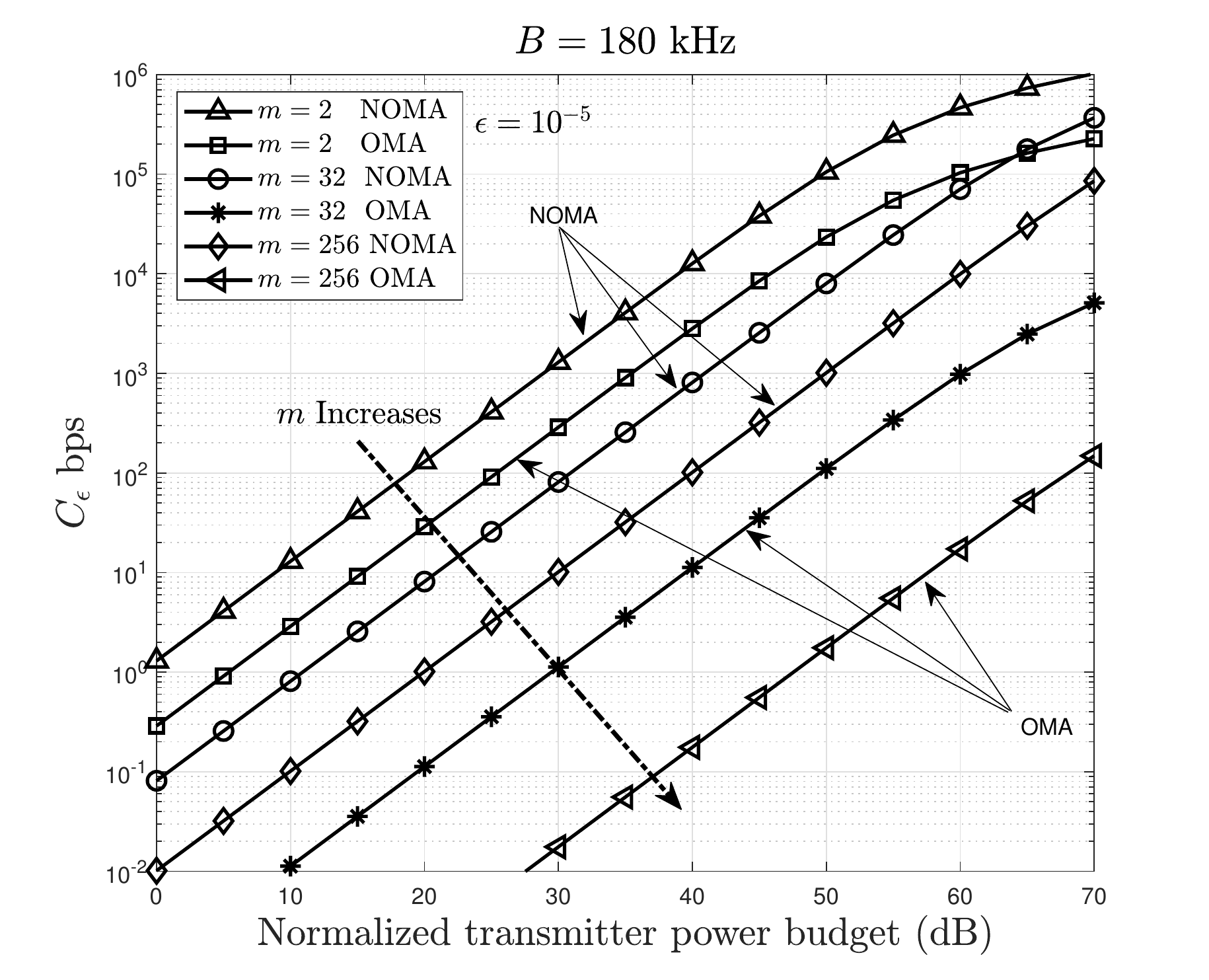}		\caption{$C_{\epsilon}$ versus normalized transmission power budget.}\label{Fig1}
	\end{figure}

In an $m$-user SISO-NOMA cluster, the transmitted signal is given by $x=\sum_{i=1}^{m}\sqrt{p_i}s_i$ such that $\sum_{i=1}^mp_i\leq p_t$, where $p_i$ is the transmitted power assigned to the $i{\text{-th}}$ NOMA user, $s_i$ is the signal to be sent to the $i{\text{-th}}$ NOMA user, and $p_t$ is the power budget of NOMA transmitter per sub-band.
Accordingly, the received signal at the $i{\text{-th}}$ user end is given by
\begin{equation}
    y_i=h_ix+w_i,~~~\forall~ i=1, \dots, m,
\end{equation}
where $h_i$ is the complex channel gain of the $i{\text{-th}}$ user and $w_i$ is the noise plus interference for the $i{\text{-th}}$ user.
For NOMA, the magnitudes of channel gains between the NOMA transmitter and receivers are first ordered  such that $|h_1|\geq \dots \geq |h_m|$. 
Thereafter, power fractions are assigned to every superimposed signal such that $p_1\leq \dots \leq p_m$. 
This is done under the condition that the received signals' power at the $i$-th NOMA user must be separated from other signals' powers by a minimum SIC hardware sensitivity value \cite{Shipon7557079}. 
Following this power allocation mechanism, the $i{\text{-th}}$ user will be able to remove all signals with higher power weights than its desired signal through multi-level SIC techniques \cite{NOMASIC2018}.
Therefore, the achievable throughput at the $i{\text{-th}}$ user is given by
\begin{equation}
    R_i=B\log_2 \left(1+\frac{p_i\gamma_i}{\sum_{j=1}^{i-1}p_j\gamma_j+1}      \right),
\end{equation}
where $B$ is the allocated bandwidth per NOMA cluster and $\gamma_i=\frac{|h_i|^2}{I_i+N_i}$, in which $I_{\text{i}}$ and $N_i$ are the inter-cluster interference (ICI) power and  the noise power at the receiver input of the $i$-th user, respectively.
To compare the performance of large-scale SISO-NOMA with that of OMA, Fig. \ref{Fig1} shows a graphical approximation of the $\epsilon$-outage capacity for SISO-NOMA with different cluster sizes ($m$). 
Here, $\epsilon$ is the maximum allowable outage to achieve a capacity of $C_{\epsilon}$.
In order to investigate SISO-NOMA at the large scale regime, we have used constant transmission power scheme that complies with NOMA protocol and does not affect the insights extracted from the provided results\footnote{Optimal transmission power allocation scheme for an $m$-user SISO-NOMA system was presented in \cite{Shipon8352618}.}.
It can be noticed that increasing the number of users per NOMA cluster causes a significant degradation on the system coding gain (i.e. the performance curve shifts to the right).

In mMTC networks, communicating devices may be significantly small in sizes and distributed within remote areas which makes it difficult to guarantee a permanent power sources for all devices at all times.
Accordingly, power consumption plays a critical rule in the design of mMTC networks and devices.
By using NOMA at the large-scale regime, higher spectral efficiency and/or lower power consumption can be achieved. 
As an example, under the same minimum rate requirements for OMA and NOMA, the required NOMA transmission power per cluster can be significantly reduced when the entire OMA bandwidth is utilized.
In general, spectrum-efficiency and energy-efficiency trade-offs can be exploited when deploying large-scale NOMA in mMTC networks.

\subsection{MIMO-Enabled NOMA}
\begin{figure}[!htb]
		\centering		\includegraphics[height=4.3cm, width=6.7cm]{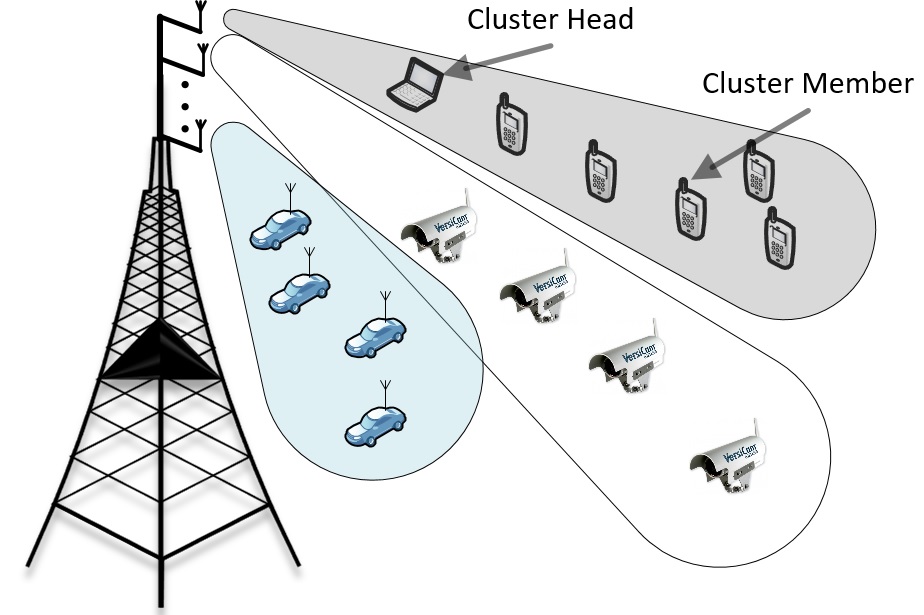}		\caption{MIMO-NOMA transmission in downlink.}\label{Fig2}
	\end{figure}

The extension of NOMA scheme to  multiple-input multiple-output (MIMO) systems can be considered as the first intuitive solution to integrate NOMA with more sophisticated wireless technologies.
There are quite a number of papers in the literature that discussed the merging between MIMO systems and NOMA scheme with different layouts.
In this section we present, as an example, a clustering algorithms proposed by \cite{Shipon8352618} to find the best utilization of MIMO-enabled NOMA system. 
Fig. \ref{Fig2} shows the proposed clustering model for MIMO-enabled downlink NOMA in which a transmitter with a group of antennas  transmits to a set of users simultaneously using the same transmission sub-band. 
Generally, every transmitting antenna defines its own cluster set of receiving users and applies NOMA-based power allocation per cluster. 
Additionally, to reduce inter-cluster interference, transmitting antennas apply beam-forming (for example, zero-forcing beamforming) to allocate every cluster a share of the overall transmission power budget \cite{Shipon7557079}.

For large-scale MIMO-enabled NOMA systems, every cluster should have an order of tens of NOMA users (or even hundreds in the case of massive in-band NOMA scheme) that share the same in-band beam (same cluster).
This increase in cluster size will result in an increase in the required transmission power per MIMO beam in order to guarantee a minimum rate per user (especially in presence of  high inter-cluster interference).
However, MIMO-enabled NOMA adds some degrees of freedom to achieve spectrum-power trade-off designs that comply with the notion of mMTC communication for which a significantly large spectrum efficiency and low power consumption are crucial.
Accordingly, using the same OMA resources (i.e. available spectrum and power budget), with the appropriate design, MIMO-enabled NOMA may achieve similar performance with relatively lower power consumption. 
\subsection{CoMP-Enabled NOMA}
In the last two subsection, we have discussed the NOMA scheme under one transmitter that has a single power budget to be allocated to different NOMA layouts (SISO and MIMO).
In this section, we briefly discuss the integration of NOMA with the so called coordinated multi-point (CoMP) transmission.
Specifically, we focus on joint transmission-CoMP (JT-CoMP) scheme in which multiple geographically distributed BSs serve different users using the same sub-band.
All of the existing literature that focused on CoMP-enabled NOMA considered dividing users into cell-edge users and cell-centre users. 
Accordingly, every NOMA cluster is assumed to consist of one or multiple of the cell edge users and a one or multiple of cell-centre users.

\begin{figure}[!htb]
		\centering		\includegraphics[height=5.0cm, width=8.5cm]{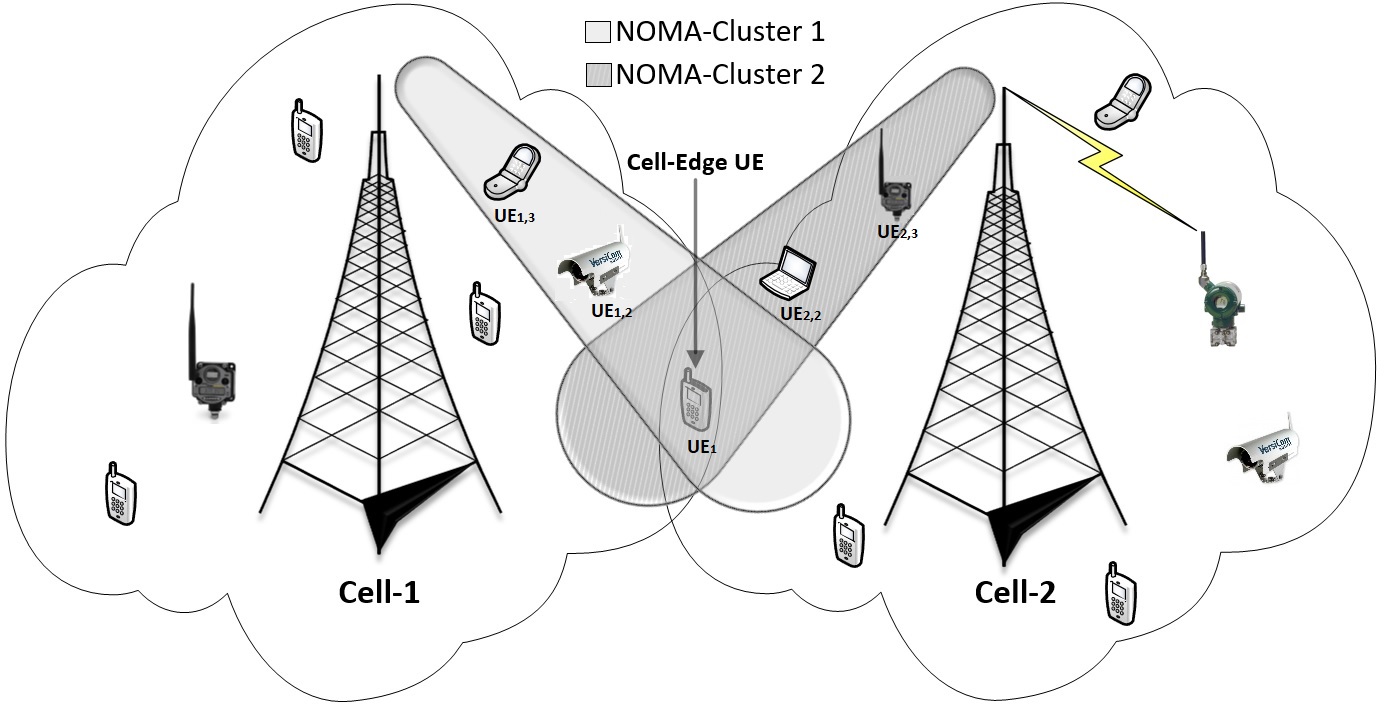}		\caption{CoMP-enabled NOMA with two cooperating BSs.}\label{Figur3}
	\end{figure}

Generally, under the JT-CoMP system, the received signal at the $i$-th user can be expressed as
\begin{equation}
   y_i=\sum_{\forall~m\in \mathbb{S}_{\text{c},i}}h_{i,m}x_m
+
\sum_{\forall~j\in \mathbb{S}_{\text{nc},i}}h_{i,j}x_j+w_i,\label{Eq.3}
\end{equation}
where $h_{i,m}$ is the complex channel gain between the $i$-th user and $m$-th BS, $\mathbb{S}_{\text{c},m}$($\mathbb{S}_{\text{nc},m}$) is the set containing the indices of BSs serving (not-serving) the $i$-th user through a single sub-band.
If the $i$-th user is identified as a cell-edge user, then more than one BS should transmit the same signal to that user.
Fig. \ref{Figur3}, shows a possible scenario of CoMP-enabled NOMA scheme.
Using this scheme, the achievable throughput at the cell-edge user (denoted $\text{UE}_1$) and at the $j$-th user served by the $i$-th BS (denoted $\text{UE}_{i,j}$) are expressed, respectively, as \cite{Shipon2018}
\begin{dmath}
R_1 = \log_2\left(1+\frac{\sum_{i=1}^2p_{i,1}\gamma_{i,1}}{\sum_{i=1}^{2}\sum_{j=2}^3p_{i,j}\gamma_{i,1}+1} \right), 
\end{dmath}
\begin{dmath}
 R_{i,j} = \log_2\left(1+\frac{p_{i,j}\gamma_{i,j}}{\sum_{k=j+1}^3p_{i,k}\gamma_{i,j}+I_{\text{inter}}+1} \right),  \label{Eq.5}
\end{dmath}
where $I_{\text{inter}}= \sum_{m=1,m\neq i}^{2}\sum_{l=2}^3p_{m,l}\gamma^'_{i,j}$ is the inter-cluster interference (ICI) caused by joint utilization of the same sub-band by two NOMA clusters.
It can be noticed from Eq. (\ref{Eq.5}) that being a cell-centre NOMA user will add an additional burden in the form of ICI ($I_{\text{inter}}$). 
In that regard, for large-scale NOMA, we have found that user performance is very sensitive to any further interfering components (in addition to INI component) even in the existence of sophisticated power allocation algorithms.
This increases the difficulty of deploying a large-scale NOMA clusters under the conventional CoMP system.
Nevertheless, CoMP-enabled NOMA with small cluster sizes was shown to significantly enhance average performance per user \cite[Fig. 3]{Shipon2018}.
\section{Limitations and Requirements for Large-Scale NOMA}
We discuss the limitations and requirements of NOMA deployment with large cluster sizes noting that studies on in-band large-scale NOMA is scanty in the existing literature.

\subsection{High Interference Levels}
Generally, there are two main interference components that affect the performance of NOMA receiver namely; the INI and the ICI.
Based on NOMA protocol, users with lower channel quality will have to deal with interference due to those with higher channel quality.
Accordingly, as the number of NOMA users per cluster ($m$) increases (see Eq. 2), users with low channel quality may suffer due to increased level of unfiltered INI components. Besides, ICI will occur due to the utilization of the same sub-band by more than one NOMA clusters. 
This can be done either within the same cell (such as MIMO-NOMA) or within different geographically separated cells (such as CoMP and multi-cell networks with a reuse factor greater than or equal to one \cite{Tabassum2018}). 
It was found that under large-scale NOMA, user performance becomes very sensitive to small values of ICI.
Therefore, sophisticated SIC mitigation technologies will be required to implement NOMA with a large cluster sizes.

\subsection{Hardware Requirements for Large-Scale NOMA}
The main idea of NOMA is based on the ability of the receiver to cancel a significant component of INI throughout SIC unit.
Generally, due to the randomness of channel gains and the mobility of NOMA wireless receivers, a certain NOMA receiver should be able to cancel up to $m-1$ signals at any arbitrary time slot (where $m$ is the in-band NOMA cluster size).
Accordingly, any design for large-scale NOMA receiver must take care of the following concerns:
\begin{enumerate}[(a)]
    \item For large-scale NOMA, SIC must be achieved quickly and efficiently since NOMA is supposed to be operating at the time slot level.
    \item The increase of NOMA receiver sensitivity to different values of ICI at large cluster sizes.
    \item At large NOMA cluster sizes, differences among the received signals' power components at a certain NOMA receiver becomes smaller, especially for limited power budget. 
    This will negatively affect the performance of SIC unit in decoding the highest weight signal.
    Accordingly, more complicated receiver designs will be required.
    \item Larger in-band NOMA cluster sizes means a larger number of required SIC operations which results in higher power consumption.
\end{enumerate}
\section{An Enabling Technique for Massive-NOMA}
\begin{figure}[!htb]
		\centering		\includegraphics[height=4.8cm, width=7.85cm]{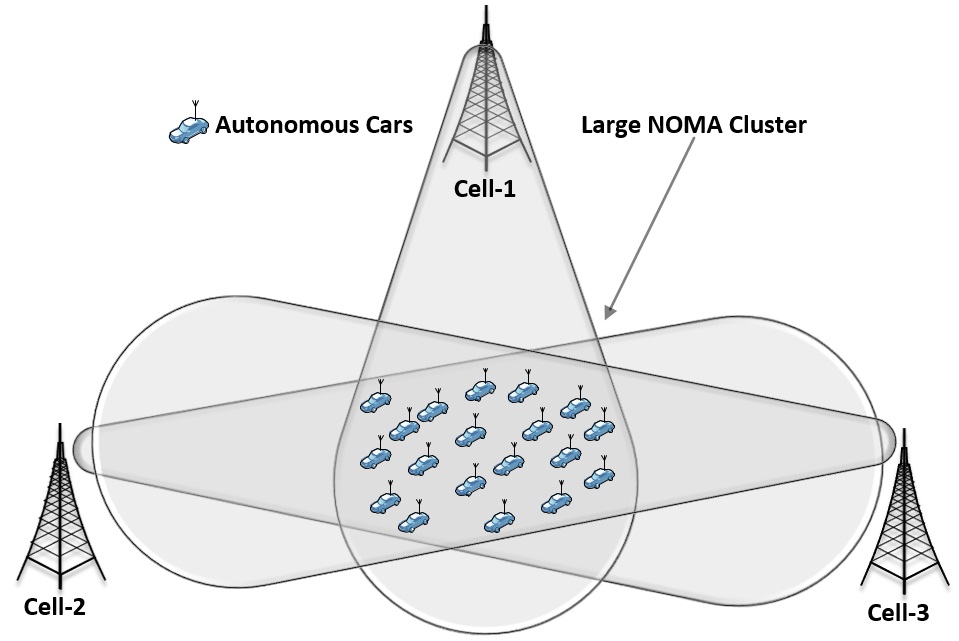}		\caption{Large-scale NOMA.}\label{Novel}
	\end{figure}

\begin{figure}[!htb]
		\centering		\includegraphics[height=5.65cm, width=7.65cm]{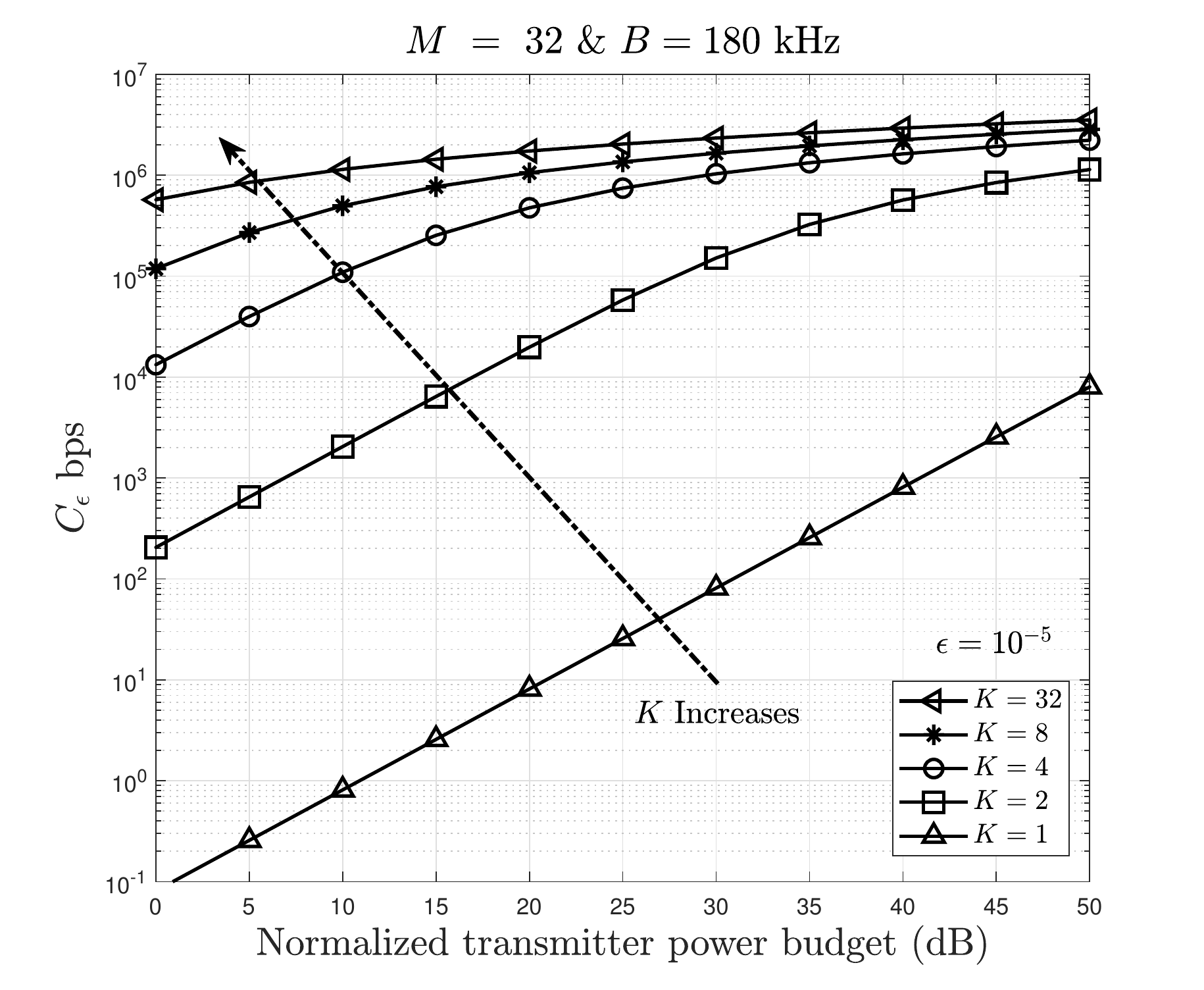}		\caption{ $C_{\epsilon}$ versus normalized transmission power.}\label{Fig6}
	\end{figure}

When NOMA signal is enhanced such that the level of the desired power at every NOMA receiver is increased, the amount of interference is decreased and complexity of SIC is reduced, then the deployment of large-scale NOMA will become feasible.
Traditional diversity enhancing techniques such as MIMO and CoMP could be a good candidate for large-scale NOMA scheme if the ICI component is mitigated efficiently.
Here, we propose a novel cooperative NOMA scheme that enables the deployment of large-scale NOMA.
Fig. \ref{Novel} shows an example of the proposed scheme at which three BSs are cooperating to serve one large-sized NOMA cluster.
Using this layout and with a proper network planing, the ICI components may be significantly reduced.
Additionally, using this system setup, the amount of desired signal at every NOMA receiver will be increased from more than one transmitter. Such an increase will result in larger power gaps (enhanced receiver sensitivity) and add more degrees of freedom on the required level of SIC for every NOMA receiver.
Fig. \ref{Fig6} shows a simulation example of the proposed scheme.
The variable $K$($M$) denotes the number of cooperating NOMA transmitters(receivers).
It can be noticed that even under significantly large $M$, increasing the number of cooperating transmitters enhances the system diversity gain significantly (increase in the slope of the performance curves).
\section{Conclusion}
NOMA can provide the required spectrum-energy efficiency trade-off required for mMTC networks. We have discussed several enabling technologies for NOMA (e.g. SISO-NOMA, MIMO-NOMA, CoMP-NOMA) and practical aspects of the deployment of large-scale NOMA (i.e. NOMA with large cluster sizes).  
To enable large-scale NOMA for massive mMTC, sophisticated diversity enhancing techniques need to be used to compensate for the severe degradation in coding gain and also to decrease the complexity of required successive interference cancellation (SIC).
\bibliographystyle{IEEEtran}
\bibliography{IEEEabrv,yasser}

\begin{IEEEbiography}[{\includegraphics[width=1in,height=1.25in,clip,keepaspectratio]{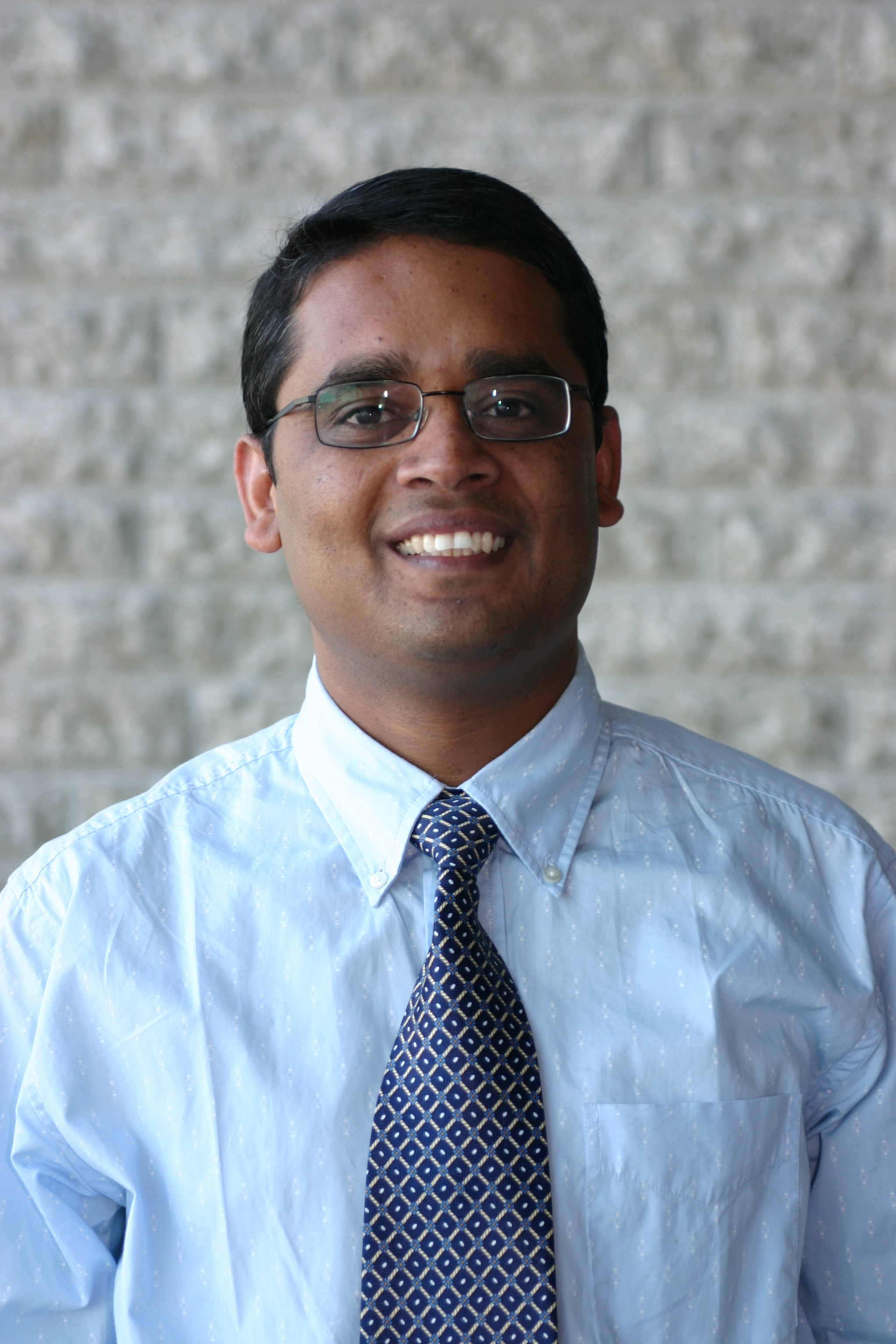}}]
{Ekram Hossain} (F'15) is a Professor in the Department of Electrical and Computer Engineering at University of Manitoba, Canada (http://home.cc.umanitoba.ca/$\sim$hossaina). He is a Member (Class of 2016) of the College of the Royal Society of Canada. Dr. Hossain's current research interests include design, analysis, and optimization of wireless, mobile, cognitive, and green communication networks with emphasis on beyond 5G cellular networks. To date, his research works have received 23,000+ citations (in Google Scholar, with h-index = 77). He was elevated to an IEEE Fellow ``for contributions to spectrum management and resource allocation in cognitive and cellular radio networks". He received the 2017 IEEE ComSoc TCGCC (Technical Committee on Green Communications \& Computing) Distinguished Technical Achievement Recognition Award ``for outstanding technical leadership and achievement in green wireless communications and networking". He was listed as a Clarivate Analytics Highly Cited Researcher in Computer Science in 2017 and  2018. Currently he serves as the Editor-in-Chief of IEEE Press. Previously he served as the Editor-in-Chief for the IEEE Communications Surveys and Tutorials (2012-2016), Area Editor for the IEEE Transactions on Wireless Communications in the area of ``Resource Management and Multiple Access'' (2009-2011), an Editor for the IEEE Transactions on Mobile Computing (2007-2012), and an Editor for the IEEE Journal on Selected Areas in Communications - Cognitive Radio Series (2011-2014). He is a Distinguished Lecturer of the IEEE Communications Society and the IEEE Vehicular Technology Society. Also, he is an elected member of the Board of Governors of the IEEE Communications Society for the term 2018-2020. 
\end{IEEEbiography}

\begin{IEEEbiography}[{\includegraphics[width=1in,height=1.25in,clip,keepaspectratio]{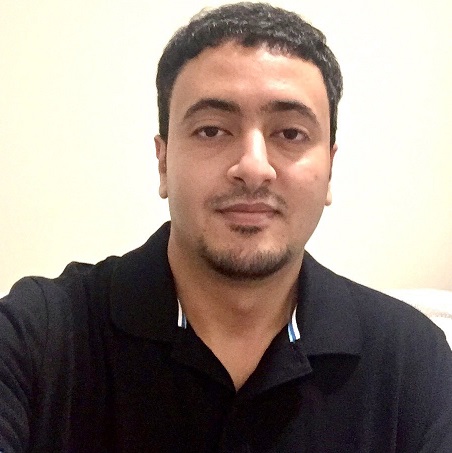}}]
{Yasser Al-Eryani} was born in Sana'a, Yemen. He received a B.Sc. degree in Electrical Engineering from IBB University, Ibb, Yemen, in 2012 and received a M.Sc. degree in telecommunications engineering from King Fahd University of Petroleum and Minerals (KFUPM), Dhahran, KSA, in 2015. 
He is now working towards his Ph.D. degree in electrical engineering at the University of Manitoba, Winnipeg, Canada. 
His research interests are design and analysis of wireless communication networks.
\end{IEEEbiography}

\end{document}